\documentclass[letterpaper,english,aps,prb,floatfix,showpacs,twocolumn,amsfonts,amssymb,superscriptaddress]{revtex4-1}

\usepackage[T1]{fontenc}
\usepackage[applemac]{inputenc}
\usepackage[english]{babel}
\usepackage{graphics}
\usepackage{amssymb}
\usepackage{amsmath}
\usepackage{overpic}

\usepackage{natbib}
\usepackage{xcolor}
\usepackage{soul}

\newcommand{\be}{\begin{equation}}
\newcommand{\ee}{\end{equation}}
\newcommand{\bea}{\begin{eqnarray}}
\newcommand{\eea}{\end{eqnarray}}

\def\a{\alpha}
\def\b{\beta}
\def\e{\varepsilon}
\def\d{\delta}
\def\g{\gamma}

\def\l{\lambda}

\def\p{\pi}
\def\s{\sigma}

\def\ra{\rightarrow}

\def\pll{\parallel}

\def\pd{\partial}

\def\bk{{\bf k}}

\def\bq{{\bf q}}
\def\br{{\bf r}}

\def\bA{{\bf A}}

\def\OO{{\cal{O}}}

\def\nn{\nonumber}
\def\lb{\label}
\def\pref#1{(\ref{#1})}

\newcount\bozza \bozza=0
\ifnum\bozza=1
\newdimen\shift \shift=-2truecm
\def\lb#1{%
{\label{#1}\rlap{\kern\shift{$\scriptstyle#1$}}}}
\else\def\lb#1{\label{#1}} \fi

\begin{document}

\title{Non-linear Terahertz Driving of Plasma Waves in Layered Cuprates}
\author{Francesco Gabriele}
\affiliation{Department of Physics and ISC-CNR, ``Sapienza'' University of Rome, P.le A. Moro 5, 00185 Rome, Italy}
\author{Mattia Udina}
\affiliation{Department of Physics and ISC-CNR, ``Sapienza'' University of Rome, P.le A. Moro 5, 00185 Rome, Italy}
\author{Lara Benfatto}
 \email{lara.benfatto@roma1.infn.it}
\affiliation{Department of Physics and ISC-CNR, ``Sapienza'' University of Rome, P.le A. Moro 5, 00185 Rome, Italy}

\begin{abstract}
The hallmark of superconductivity is the rigidity of the quantum-mechanical phase of electrons, responsible for superfluid behavior and Meissner effect.  The strength of the phase stiffness is set by the Josephson coupling, which is strongly anisotropic in layered superconducting cuprates.  So far, THz light pulses have been efficiently used to achieve non-linear control of the out-of-plane Josephson plasma mode, whose frequency scale lies in the THz range. However, the high-energy in-plane plasma mode has been assumed to be insensitive to THz pumping. Here, we show that THz driving of both low-frequency and high-frequency plasma waves is possible via a general two-plasmon excitation mechanism. The anisotropy of the Josephson couplings leads to marked differences in the thermal effects among the out-of-plane and in-plane response, consistently with the experiments. Our results link the observed survival  of the in-plane THz non-linear driving above $T_c$ to enhanced fluctuating effects in the phase stiffness in cuprates, paving the way to THz impulsive control of phase rigidity in unconventional superconductors. 

\end{abstract}
\date{\today}

\maketitle

Order and rigidity are the essential ingredients of any phase transition. In a superconductor the order is connected to the amplitude of the complex order parameter, related to the opening of a gap $\Delta$  in the single-particle excitation spectrum. The rigidity manifests instead in the quantum-mechanical phase of the electronic wave function, associated with the phase of the order parameter\cite{nagaosa}.  Twisting the phase is equivalent to an elastic deformation in a solid, meaning that its energetic cost is vanishing for sufficiently slow spatial variations.  On the other hand, since phase fluctuations come along with charge fluctuations, long-range Coulomb forces push the energetic cost of a phase gradient to the plasma energy $\omega_J$\cite{anderson_pr58,nagaosa}. While for ordinary superconductors this energy scale is far above the THz range, in layered cuprates 
the weak Josephson coupling among neighboring layers\cite{shibauchi_prl94,panagopoulos_prb96,bonn_prl04} pushes down the frequency of the inter-layer Josephson plasma mode (JPM) to the THz range.\cite{nori_review10,cavalleri_review} The possibility to manipulate the inter-layer JPM by intense THz pulses has been theoretically discussed long ago within the context of the non-linear equation of motion\cite{koyama_prb96,machida_prl99,machida_physc00,nori_natphys06,nori_review10,cavalleri_review}. This approached turned out to successfully capture the main features of a  series of recent experiments\cite{cavalleri_natphys16,cavalleri_science18}, even though a full quantum treatment of the JPM able to capture thermal effects across $T_c$ is still lacking. On the other hand, non-linear effects induced by strong THz pulses polarized in the planes\cite{shimano_prl18,kaiser_natcomm20,shimano_cm19}  have been discussed so far only within the context of the SC amplitude (Higgs) mode, whose excitation energy $\omega_H=2\Delta$, ranging from 5 to 10 THz in cuprates, appears as  a better candidate than high-energy in-plane plasma waves. Nonetheless,  
the observed monotonic temperature dependence of the non-linear response\cite{kaiser_natcomm20,shimano_cm19}, its persistence above $T_c$\cite{shimano_cm19} and its polarization dependence\cite{shimano_prl18} do not easily match the expectations for the Higgs mode. The same problem holds considering lattice-modulated charge fluctuations,  which are expected to dominate in the clean limit\cite{cea_prb16,udina_prb19}  but become less relevant\cite{silaev_prb19,shimano_prb19,shimano_review19,seibold_cm20} and strongly isotropic\cite{seibold_cm20} when disorder is considered. 
%As it has been discussed within the context of conventional superconductors Since this is the same scale where  collective density fluctuations proliferate in a superconductor, a resonant response at $2\Delta$ in general cannot be uniquely linked to the Higgs mode\cite{cea_prb16,shimano_prb17,cea_prb18,shimano_review19,silaev_prb19,udina_prb19}. On the other hand, in cuprates both the temperature dependence of the third-harmonic generation (THG) and its polarization dependence of the signal do not match neither of the two paradigms for the Higgs or charge fluctuations, leaving open the question of the presence of an additional contribution to the SC non-linear response. 
%
\begin{figure}
\centering
\includegraphics[scale=0.6]{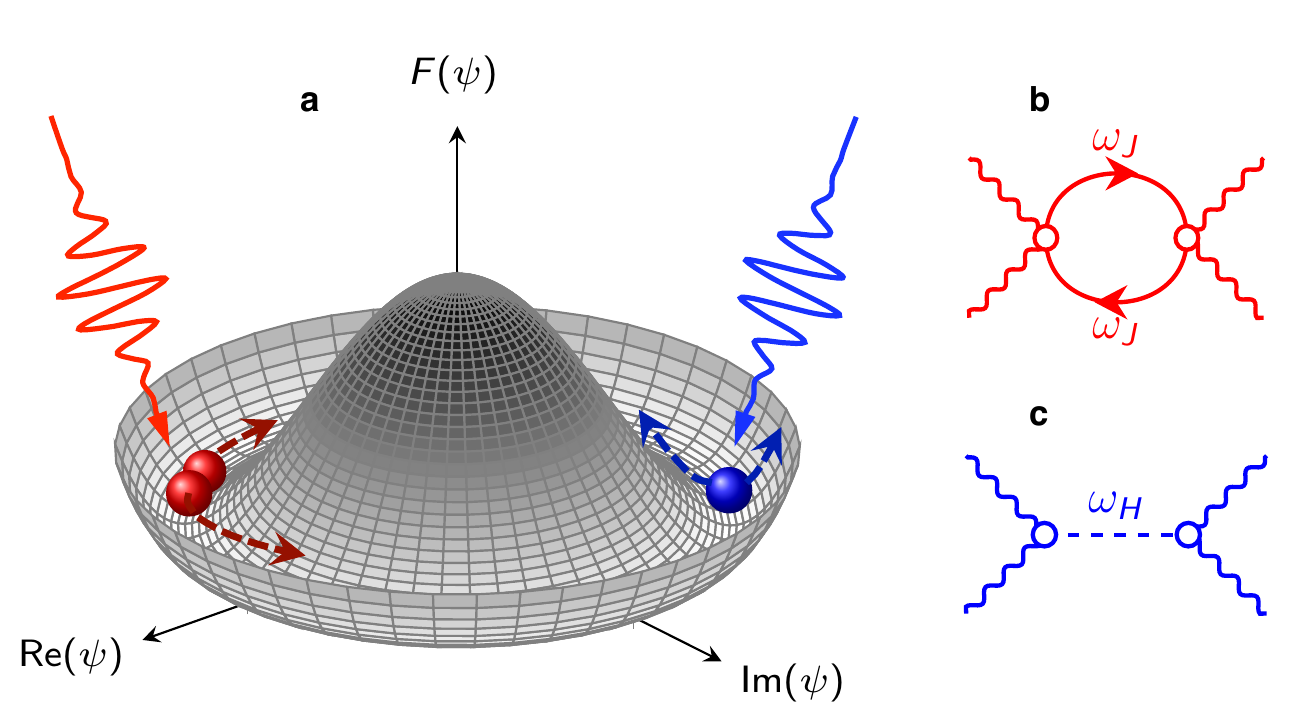}
\caption{(a) Schematic view of the mexican-hat potential for the free energy $F(\psi)$, with $\psi$ the complex order parameter of a superconductor below $T_c$. A phase-gradient excitation corresponds to a shift along the minima, while a Higgs excitation moves the system away from the minimun. An intense light pulse with almost zero momentum can excite simultaneously two plasma waves with opposite momenta (in red) or a single Higgs fluctuation (in blue). (b)-(c) Feynman-diagrams representation of the (b) plasma-waves or (c) Higgs contribution to the non-linear optical response. Here wavy lines represent the e.m.\ field, solid/dashed lines the plasmon/Higgs field, respectively.}
\label{fig1}
\end{figure}
%\cite{shimano_prl12,shimano_prl13,shimano_science14,wang_natphot2019,shimano_review19}. 

Here  we provide a complete theoretical description of the JPM contribution to the non-linear response of layered cuprate superconductors, focusing both on third-harmonic generation (THG) and pump-probe protocols. We first show that the basic mechanism behind non-linear photonic of Josephson plasma waves is intrinsically different from the one of the Higgs mode, see Fig.\ \ref{fig1}. By pursuing the analogy with lattice vibrations in a solid, the Higgs mode is like a Raman-active optical phonon mode. It has a finite frequency at zero momentum, and its symmetry allows for a finite quadratic coupling to light\cite{shimano_science14,aoki_prb15,cea_prb16,silaev_prb19,shimano_prb19,shimano_review19,udina_prb19,manske_natcomm20,wang_natphot2019,seibold_cm20}. The phase mode behaves instead like an acoustic phonon mode, pushed to the plasma energy by Coulomb interaction, carrying out a finite momentum at nonzero frequency. As such, zero-momentum light pulses can only excite simultaneously two JPMs with opposite momenta, making this process strongly dependent on the thermal probability to populate excited states. This feature differentiates drastically the temperature dependence of the THG associated with out-of-plane or in-plane JPMs, since the frequency scale of the former is comparable to $T_c$, while it is much larger for the latter. In addition, in contrast to the Higgs mode\cite{cea_prb16,shimano_review19}, for a light pulse polarized in the planes the signal coming from JPMs is in general anisotropic, since the momenta carried out by the two plasmons can be along different crystallographic axes. All these features not only contribute to the understanding of the existing experimental measurements,\cite{cavalleri_natphys16,cavalleri_science18,shimano_prl18,kaiser_natcomm20,shimano_cm19} but they also offer a perspective to design future experiments aimed at selectively tune non-linear photonic of Josephson plasma waves in layered cuprates.

Let us first focus on the out-of-plane JPM. We take a layered model with planes stacked along $z$. In the SC state the Josephson coupling $J_\perp$ of the SC phase $\phi_n$ between neighboring planes sets an effective XY model:
\be
\lb{hxy}
H=-J_\perp\sum_n\cos(\phi_n-\phi_{n+1}).
\ee
 An electric field polarized along $z$ enters the Hamiltonian via the minimal-coupling substitution\cite{nagaosa} $\theta_n\ra \theta_n-(2\pi/\Phi_0)d A_z$, with  $\theta_n=\phi_n-\phi_{n+1}$ , $d$ interlayer distance and $\Phi_0=hc/(2e)$. The corresponding out-of-plane current density $I_z=-\pd H/\pd (c A_z)$ is given by:
\be
\lb{curr}
I_z=J_c \sin (\phi_n-\phi_{n+1}-(2\pi/\Phi_0)d A_z),
\ee
where $J_c=2e J_\perp/\hbar S$, with $S$ surface of each plane. The Josephson current \pref{curr} naturally admits an expansion in powers of $A_z$ to all orders:
\be
\lb{nonl}
\langle I_z \rangle= \chi_z^{(1)} A_z +\chi^{(3)}_{z} A_z^3+\cdots,
\ee
where the explicit time convolution of Eq.\ \pref{nonl} has been omitted for compactness. Here, following the same approach used so far to investigate the Higgs response\cite{shimano_science14,aoki_prb15,cea_prb16,udina_prb19}, we rely on a quasi-equilibrium description, where the leading effect of the intense THz pump field is to trigger a third-order $\chi^{(3)}$ response mediated by plasma waves. The quantum generalization of the model \pref{hxy} has been widely discussed within several contexts\cite{machida_prl99,machida_physc00,benfatto_prb01,nori_review10,benfatto_prb04}. Here we follow the approach of Ref.s \cite{benfatto_prb01,benfatto_prb04} where long-range Coulomb interactions are introduced within a layered model appropriate for cuprates (see Ref.\cite{suppl}). 
%Following the microscopic approach developed in Refs.\ \cite{machida_prl99,machida_physc00,nori_review10}, 
The Gaussian quantum action for the phase mode at long wavelength has the usual form:
\bea
\lb{sgaussapp}
S^G_\perp&\simeq& \frac{vS}{2d}\sum_{i\omega_m,k_z} 4\sin^2(k_zd/2) \left[ \omega_m^2+\omega_J^2\right]|\phi(i\omega_m,k_z)|^2, \hspace{0.5cm}
\eea
where $\omega_J^2=c^2/\e \l_c^2={8\pi e d J_c/\hbar \e}$ is the energy scale of the out-of-plane JPM,  $i\omega_m=2\pi m T$ are Matsubara frequencies and $v=\hbar^2\e/(16\pi e^2)$, with $\e$ the background dielectric constant. In the classical limit  only $\omega_m=0$ is relevant and one recovers the leading term of Eq.\ \pref{hxy}, i.e. a discrete phase gradient along $z$, as expected for the Goldstone mode. 
%The finite-momentum dispersion of the plasma mode, which is crucial to describe propagating waves within the context of the linearized sine-Gordon equation\cite{machida_prl99,machida_physc00,nori_review10,cavalleri_natphys16,cavalleri_science18}, is quantitatively irrelevant for the behavior of the non-linear optical kernel\cite{suppl},  so we will focus in what follows on the model \pref{sgaussapp}.  
 
To compute the third-order contribution in Eq.\ \pref{nonl} we need to derive the effective action $S_\bA^{(4)}$ for the gauge field up to terms of order $\OO(A_z^4)$ (see Ref.\ \cite{suppl}). By coupling the gauge field $A_z$ to the phase mode via the minimal-coupling substitution in Eq.\ \pref{curr} and by expanding the cosine term, one finds that:
\be
\lb{stot}
S=S^G_\perp-\frac{\pi^2 d^2}{\Phi_0^2} \sum_{n,\br_i}J_\perp\int d\tau A_z^2( \tau)\theta^2_{n,\br_i}(\tau)+\cdots,
\ee
where dots denote additional terms not relevant for the $\chi^{(3)}$ response. The second term in Eq.\ \pref{stot} can be treated as a perturbation with respect to $S_G$\cite{suppl}, so that integrating out the JPM one obtains:
\bea
S_\bA^{(4)}&=&\int d t \int d t' A_z^2( t)K_\perp(t-t')A_z^2(t')=\nn\\
\lb{s4}
&=&\int d\omega A_z^2(\omega)K_\perp(\omega)A_z^2(-\omega)
\eea
where $A_z^2(\omega)$ is defined as the Fourier transform of $A_z^2(t)$. $K_\perp(\omega)$ is the non-linear optical kernel of the system, given by the convolution of two JPM propagators, as represented diagrammatically in Fig.\ \ref{fig1}b: 
\be
\lb{knonl}
K_\perp(\omega)=K_0 \frac{J_\perp^2}{\omega_J}\frac{\coth(\beta \omega_J/2)}{4\omega_J^2-(\omega+i\gamma)^2},
\ee
with $K_0$ a constant prefactor and $\gamma$ accounting for the plasmon dissipation.\cite{suppl} From Eq.\ \pref{s4} it immediately follows that $\langle I_z^{NL}(t)\rangle =\int dt' A_z(t)K(t-t')A_z^2(t')$. Therefore, for a monochromatic incident field $A_z=A_0 \cos(\omega t)$ the non-linear current admits a term oscillating at $3\omega$, whose intensity is given by\cite{shimano_science14,aoki_prb15,cea_prb16,udina_prb19}
\be
\lb{ithg}
I^{THG}=I_0 |K(2\omega)|^2,
\ee
where $I_0$ is an overall constant.  The vanishing of the denominator in Eq.\ \pref{knonl} identifies the resonance of the non-linear kernel. Since the physical mechanism behind the THG is the excitation of two plasma waves, the largest $I^{THG}$ in Eq.\ \pref{ithg} occurs when the pump frequency matches the plasma frequency, i.e.\ $\omega=\omega_J$. This has to be contrasted e.g.\ to the case of the THG from the Higgs mode. In this case the e.m.\ field excites non-linearly a {\em single} amplitude fluctuation $\delta \Delta$,  via a term like $A^2\delta \Delta$.\cite{shimano_science14,aoki_prb15,cea_prb16,udina_prb19} As a consequence the non-linear kernel, identified by the dashed line in Fig.\ \ref{fig1}c, is proportional to a single Higgs fluctuation, and the THG \pref{ithg} is resonant when the pump frequency matches {\em half} the mode energy, i.e. $\omega=\omega_H/2=\Delta$, as observed in conventional superconductors\cite{shimano_science14,wang_natphot2019}. 

\begin{figure*}
\centering
\includegraphics[scale=0.6]{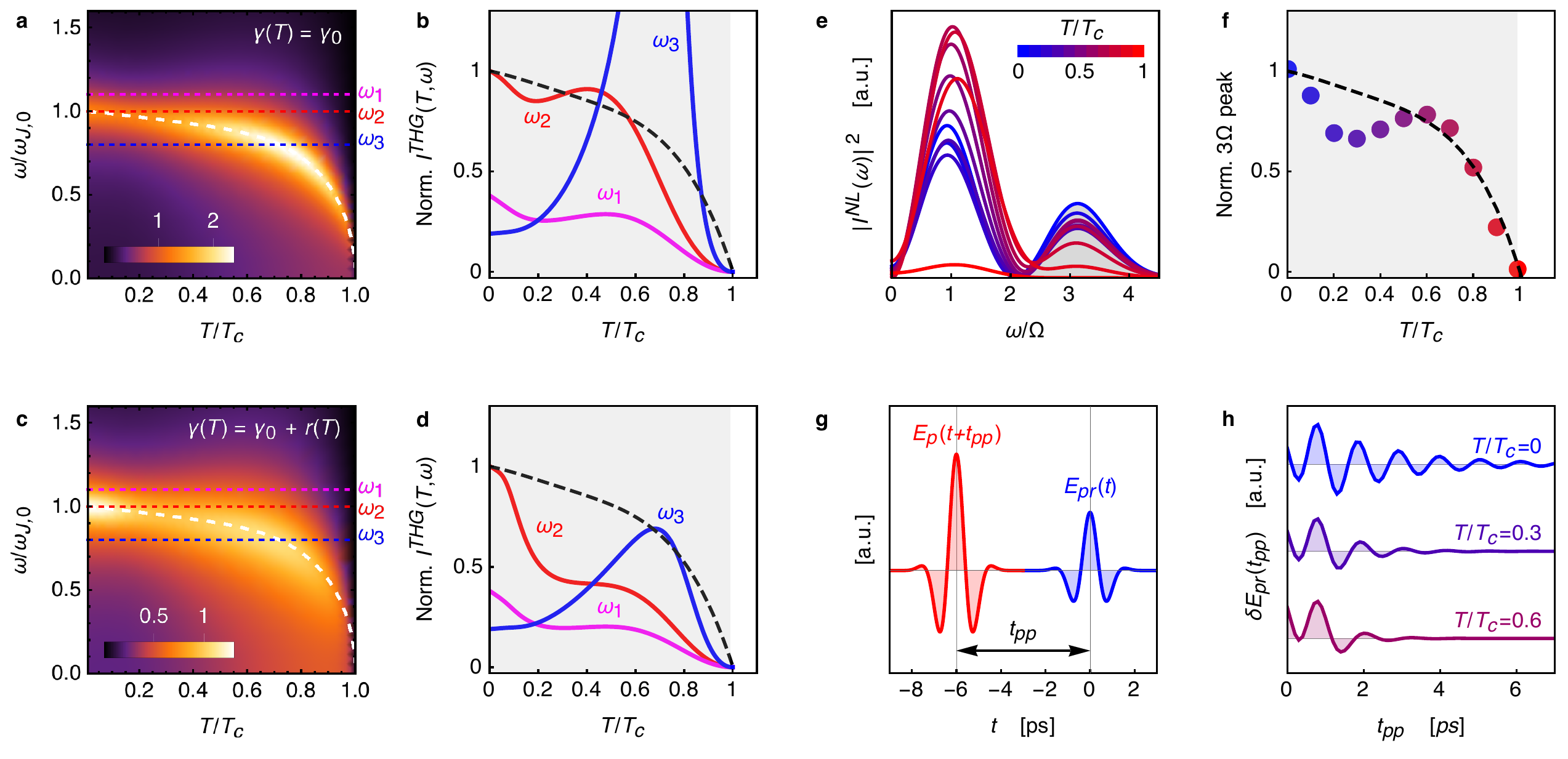}
\caption{ (a)-(d) Narrow-band pulse. Temperature and frequency dependence of the non-linear kernel \pref{knonl} for constant (a) and temperature-varying (c) damping $\g$. The dashed line denotes $\omega_J(T)/\omega_{J,0}$. Panels (b),(d) show the corresponding $I^{THG}(\omega_i,T)$ for three values $\omega_i$ of the pump frequency, normalized to $I^{THG}(\omega_{J,0},0)$. The dashed line represents $J_\perp(T)/J_\perp(0)$. (e)-(h) Broad-band pulse. (e) Spectrum of the non-linear current  $I_z^{NL}$ as a function of frequency, normalized to the central frequency $\Omega$ of the pump pulse, shown explicitly in panel (g). The THG signal is now obtained by integrating the peak around $3\Omega$ (gray region in panel (e)). Its temperature dependence is shown in panel (f), with the same color code of the curves of panel (e). (g) Schematic of the pump-probe set-up: a weak probe field impinges on the sample with a variable time delay $t_{pp}$ with respect to the intense pump pulse. (g) Time-dependence of the differential probe field $\delta E(t_{pp})$ measured with and without the pump, at different temperatures. The periodicity of the oscillations matches the $2\omega_J(T)$ value at each temperature. Here we set $\omega_{J,0}/2\pi=0.47$ THz in accordance to the experiments\cite{cavalleri_natphys16}. }
\label{fig2}
\end{figure*}

The temperature dependence of the JPM non-linear kernel \pref{knonl} and the corresponding THG \pref{ithg} for a narrow-band pulse are shown in Fig.\ \ref{fig2}a-d for different values of the pump frequency $\omega$. Here we modelled $J_{\perp}(T)$ and the corresponding $\omega_J(T)$ according to the out-of-plane superfluid stiffness measured in Ref.\ \cite{cavalleri_science18}. As we explained,  the possibility to match the resonance condition in the THG \pref{ithg} depends on the relative value of the pump frequency $\omega$ with respect to $\omega_J(T=0)\equiv \omega_{J,0}$. In the case where $\omega<\omega_{J,0}$, as for $\omega=\omega_3$ in Fig.\ \ref{fig2}a, the temperature-dependence of $I^{THG}(\omega_3)$ is dominated by the maximum at the temperature where $\omega_J(T)=\omega_3$. On the other hand, when $\omega\geq \omega_{J,0}$, as it is the case for $\omega=\omega_1,\omega_2$, the resonant excitation of the plasma mode cannot occur. However, $I^{THG}$ is still non-monotonic, see Fig.\ \ref{fig2}b, due to the fact that by increasing temperature the $J^2_\perp/\omega_J(T)\propto \omega^3_J(T)$ prefactor decreases, while the $\coth(\beta \omega_J(T)/2)$ term increases, accounting for the thermal excitation of plasma modes. This thermal effect is particularly pronounced for the out-of-plane JPM since $\omega_{J,0}$ is of the same order of the critical temperature $T_c$. The absolute value of $I^{THG}$ depends also on the damping $\gamma$ present in Eq.\ \pref{knonl}, which plays the same role\cite{suppl} of a linear damping term in the equations-of-motion approach. In Fig.\ \ref{fig2}c,d we show the results for a temperature-dependent $\g(T)=\g_0+r(T)$, where $r(T)=r_0e^{-\Delta/T}$ has been taken in analogy with previous work\cite{nori_natphys06} to mimics dissipative effects from normal quasiparticles. In this case the plasma resonance is progressively smeared out by increasing temperature, and for out-of-resonance conditions the THG signal looses rapidly intensity  as the system is warmed up.
%\bea
%S&=&\sum_n \int d\br d\tau \left\{ \frac{C_0}{2}\left(\frac{\pd \tilde \phi_n}{\pd \tau}+V_n\right)^2
%+\frac{\tilde J_\perp}{2\SS} \cos(\phi_n-\phi_{n+1})+\right.\nn\\
%&+&\frac{\tilde J_\pll}{2}\left(\nb \theta_n-\bA^\pll_n\right)^2+\frac{C}{2}(V_n-V_{n+1})^2+\frac{C}{2\e}(\bA^\pll-\bA^\pll_{n+1})^2
%\eea

The THG for a field polarized along $z$ has been measured so far only by means of a broadband pump.\cite{cavalleri_science18} To make a closer connection with this experimental setup we then simulated (see Ref.\ \cite{suppl}) the THG for a short ($\tau = 0.85$ ps) pump pulse $E_p(t)$ with central frequency $\Omega/2\pi=0.45$ THz, as shown in Fig.\ \ref{fig2}g. The frequency spectrum of the resulting non-linear current $I_z^{NL}$ presents then a broad peak around $3\Omega$, as shown in Fig.\ \ref{fig2}e. The integrated spectral weight of the $3\Omega$ peak is shown in Fig.\ \ref{fig2}f at several temperatures. Following Ref.\ \cite{cavalleri_science18} we used $\Omega\simeq \omega_{J,0}$, so the narrow-band response should corresponds to the case $
\omega=\omega_2$ of Fig.\ \ref{fig2}d. However, the broadband spectrum of the pump pulse enhances the response at intermediate temperatures and apart from a small deep around $T=0.2T_c$ the signal scales with the superfluid stiffness, in good agreement with the available experimental data. In the broad-band case the nature of the non-linear kernel can also be probed via a typical pump-probe experimental setup, schematically summarized in Fig.\ \ref{fig2}g. As it has been theoretically described in Ref.\ \cite{giorgianni_natphys19,udina_prb19} for the transmission geometry, the oscillations of the differential probe field with and without the pump $\delta E_{pr}(t_{pp})$ as a function of the pump-probe time delay can be directly linked to the resonant non-linear optical kernel. In the case of the out-of-plane response \pref{knonl} one then obtains (see Ref.\ \cite{suppl}): 
\bea
& &\delta E_{pr}(t_{pp}) \propto \int dt  K_\perp(t_{pp}-t)A^2_{z}(t)\nn\\
\lb{osc}
&=&F(T)\int_{-\infty}^{t_{pp}} e^{-\gamma (t-t_{pp})}\sin(2\omega_J (t_{pp}-t))A^2_{z}(t)
\eea
where $F(T) \equiv {J_\perp^2 \coth(\beta \omega_J/2)}/{\omega_J^2}$. When the pump pulse is short enough one can approximate $A^2_{z}(t) \simeq \delta(t)$ and Eq.\ \pref{osc} shows that the differential field $\delta E_{pr}(t_{pp})$ oscillates at {\em twice} the JPM frequency, and not at the frequency of the mode, as it occurs for the Higgs mode observed in conventional superconductors\cite{shimano_prl13}. This prediction is confirmed when a realistic pump pulse is used in Eq.\ \pref{osc}, as shown in Fig.\ \ref{fig2}h, which reproduces very well the $2\omega_J$ oscillations reported at low-temperature in pump-probe experiments in reflection geometry\cite{cavalleri_natphys16}. 
%
%
%As one can see, the oscillations of the differential probe field clearly match the $2\o_J(T)$ value at each temperature, and they are progressively damped at higher temperatures due to the increase of the plasmon damping $\gamma(T)$. Our results 
\begin{figure}
\centering
\includegraphics[scale=0.6]{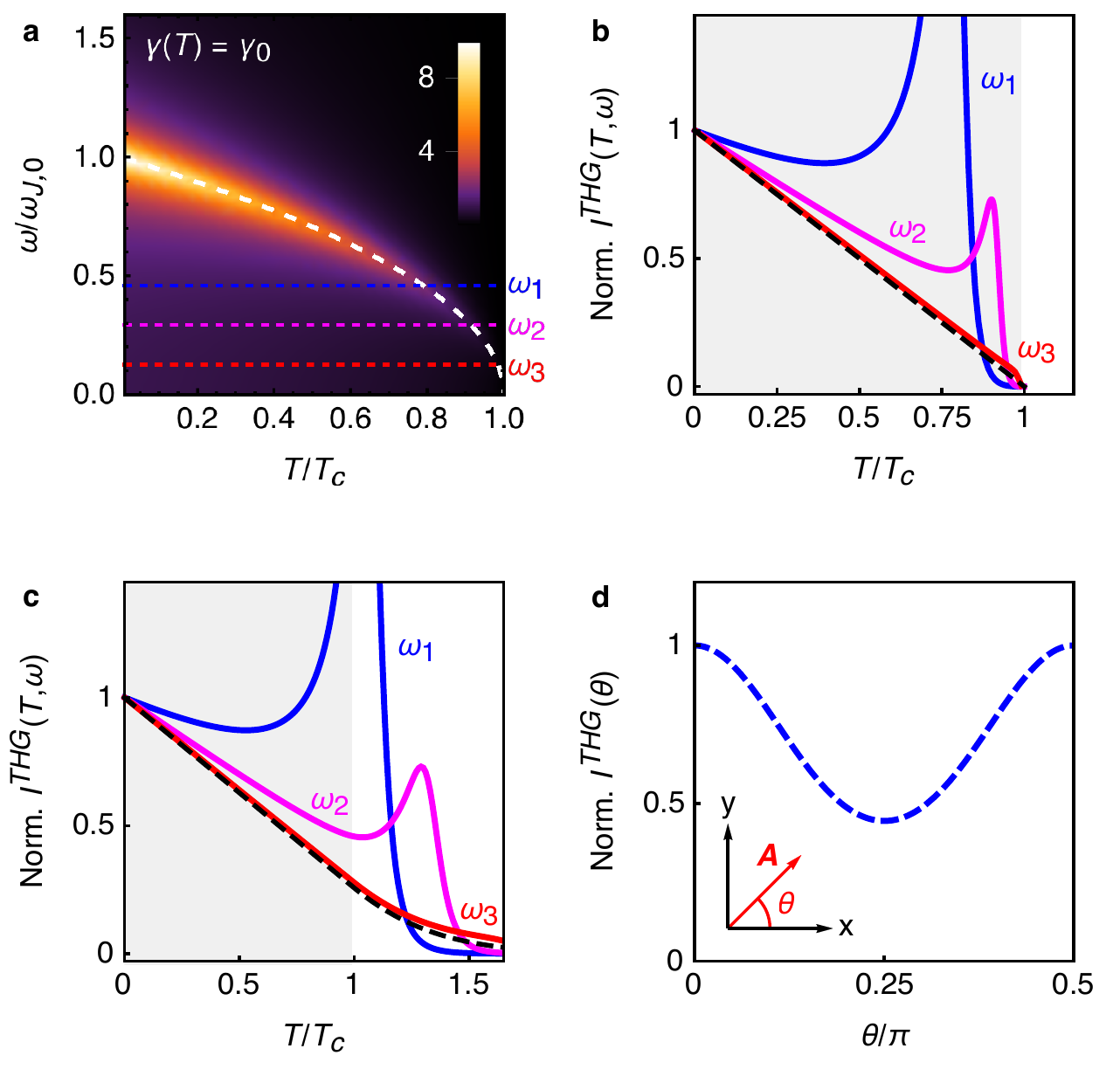}
\caption{(a) Temperature and frequency dependence of the non-linear kernel $K_\pll$ for constant damping $\g$. The dashed line denotes $\omega^\pll_J(T)/\omega^\pll_{J,0}$. (b)  $I^{THG}(\omega_i,T)$ for three $\omega_i$ values of the pump frequency, marked in panel (a), normalized to $I^{THG}(\omega_i,0)$. The dashed line represents $J_\pll(T)/J_\pll(0)$. As one can see, when $\omega^\pll_{J,0}/\omega_i$ is increased the THG intensity progressively approaches the temperature dependence of the stiffness. (c) Effect of superconducting fluctuations on the THG. Here the dashed lines simulates the experimental behavior of the $J_\pll(T)$ measured at THz frequencies, with a pronounced tail above $T_c$. When $\omega^\pll_{J,0}/\omega_i$ is increased  also the THG signal survives above $T_c$, following the fluctuating stiffness. (d) Angular dependence of $I^{THG}(\theta)=|K(\theta)|^2$, where $K(\theta)$ is given by Eq.\ \pref{ktheta}.}
\label{fig3}
\end{figure}

Let us consider now the effects of a strong THz pulse polarized within the plane.
%CuO plane of cuprates. As mentioned in the \textcolor{red}{Introduction}, the general expectation so far has been that the in-plane plasma modes in cuprates play no role in the non-linear optical response. The reason is that the in-plane plasma frequency $\o^\pll_J$ is usually of the order of 1 eV, so it cannot be resonantly excited at low temperatures by conventional pump fields with central frequencies of the order of few THz. On the other hand, the larger value of $\o^\pll_J$ follows also from a larger value of the in-plane stiffness $J_\pll$, so we expect a much larger prefactor in the non-linear coupling between the in-plane gauge field $A_\pll$ and the phase fluctuations. In addition, when the field is applied in the plane one cannot neglect in general the contribution of the charge fluctuations and the Higgs mode, which are instead absent for the out-of-plane response since the low inter-layer hopping generates incoherent out-of-plane transport. However, experiments performed so far in cuprates with an in-plane field $A_\pll$ show a rather puzzling behavior, which does not fit the theoretical predictions neither for the Higgs mode nor for the charge fluctuations. It is then worth exploring the consequence of the two-JPM process active for out-of-plane pumping in the case on a pump polarized along the SC planes, without trying a quantitative relative comparison among the other known contributions. 
In this case we can generalize the model \pref{sgaussapp} by taking into account both the two-dimensional nature of the phase fluctuations in the plane and the anisotropy of penetration depth measured experimentally in cuprates,\cite{shibauchi_prl94,panagopoulos_prb96,bonn_prl04} where $\l_c\simeq 10-100 \lambda_{ac}$ depending on the material and the doping, and $\l_{ac}\simeq 2000$ \AA, so that $\omega_J^\pll=c/\sqrt{\e} \lambda_{ac} $ is much larger than the out-of-plane one. Following again the microscopic derivation outlined e.g.\ in Ref. \cite{benfatto_prb01,benfatto_prb04} we obtain
%  Since in this case the coupling between planes is not crucial To simplify the calculations, we will consider a purely two-dimensional model and we will use the Gaussian quantum action microscopically derived e.g.\ in Ref.\ \cite{benfatto_prb04}. The structure is fully equivalent to Eq.\ \pref{sgaussapp}, with the only difference that in-plane phase fluctuations can occur along the two $(x,y)$ in-plane direction, so that $\bk=(k_x,k_y)$ and 
%
\bea
\lb{sgaussin}
S^G_\pll&\simeq& \frac{v}{2}\sum_{i\omega_m,\bk} \bk^2  \left[ \omega_m^2+(\omega^\pll_J)^2\right]|\phi(i\omega_m,\bk)|^2,
\eea
where  $\bk=(k_x,k_y)$ and we promoted the phase difference to a continuum gradient for the in-plane phase mode. To describe the non-linear coupling to the e.m.\ field we rely again on a  quantum $XY$ model,  whose coupling constant is the effective in-plane stiffness $J_\pll=\hbar^2c^2d/16\pi e^2\lambda_\pll^2$. Even though the microscopically-derived phase-only action is not in general equivalent to the $XY$ model\cite{benfatto_prb04}, for cuprates this can still represents a reasonable starting point\cite{benfatto_prb01}. By minimal-coupling substitution $\nabla \phi(\br)- (2\pi/\Phi_0) \bA_\pll$ we then obtain, in full analogy with Eq.\ \pref{stot},  that:
\be
\lb{stotin}
S=S^G_\pll-\frac{\pi^2 J_\pll}{\Phi_0^2} \int d\br d\tau \left[A_x^2( \tau)(\pd_x \phi)^2+ A_y^2( \tau)(\pd_y \phi)^2\right]+\cdots.
\ee
By following the same steps  as before we obtain a quartic action of the form \pref{s4}, but the non-linear kernel becomes a tensor which admits two different $K_{xx;xx}$ and $K_{xx;yy}$ components (see Ref.\ \cite{suppl}): 
\be
\lb{ktens}
K_{xx;xx}=3 K_\pll, \quad K_{xx;yy}=  K_\pll
\ee
where $K_\pll$ has the same structure of Eq.\ \pref{knonl}, provided that $J_\perp$ and $\omega_J$ are replaced by $J_\pll$ and $\omega_J^\pll$. The frequency and temperature dependence of $K_\pll$ is shown in Fig.\ \ref{fig3}a. The in-plane stiffness $J_\pll$  is taken as linearly decreasing, in analogy with experiments\cite{shibauchi_prl94,panagopoulos_prb96,bonn_prl04}. Since  $\omega_J^\pll(T=0)\equiv \omega^\pll_{J,0}$ is of the order of the eV, we only considered the case of THz pump frequencies $\omega_i<\omega_{J,0}^\pll$. As one can see, when $\omega_i$ is  a fraction of $\omega_{J,0}$ the resonance condition $\omega_i=\omega_J^\pll(T)$  is still attained at temperatures where the kernel is large enough to give rise to a pronounced maximum in the THG intensity. However, when $\omega_i\ll \omega_{J,0}^\pll$ the resonance is only attained near to $T_c$ where the prefactor has already washed out the two-plasmon resonace, and the THG scales with the superfluid stiffness. This can be easily seen from Eq.\ \pref{knonl}, since by putting $\omega \simeq 0$ in the denominator, and considering that $\coth (\b \omega^\pll_J)\simeq 1$ at all relevant temperatures, from $\omega_J^\pll \propto \sqrt{J_\pll}$ one finds
\be
\lb{simple}
I^{THG}(T, \omega \ll \omega^\pll_{J,0})\sim J^\pll(T).
\ee
The scaling of the THG intensity in the THz regime with $J_\pll$ has several consequences. First, $I^{THG}$ monotonically increases below $T_c$, in striking contrast with the pronounced maximum one would expect for a resonance at $\omega_i=2\Delta(T)$, due to the Higgs\cite{shimano_science14,aoki_prb15} or charge fluctuations\cite{cea_prb16,udina_prb19}. Second, the superfluid stiffness appearing in the THG response is the one measured at THz frequencies. As such, due to pronounced fluctuations effects at this frequency scale, it vanishes in cuprates well above $T_c$\cite{corson_nature99,armitage_natphys11,shimano_cm19}. The $I^{THG}$ at pump frequencies significantly smaller than $\omega^\pll_{J,0}$ closely follows the same behavior, as we exemplify in Fig.\ \ref{fig3}c where we report a simulation of the superfluid stiffness with a fluctuation tail above $T_c$.  Interestingly, both the monotonic suppression\cite{kaiser_natcomm20} and the persistence of non-linear effects above $T_c$\cite{kaiser_natcomm20,shimano_cm19} have been recently reported in THG and THz Kerr measurements in cuprate superconductors. Finally,  due to the tensor structure of the in-plane kernel  \pref{ktens}, the non-linear current associated with JPM for a pump field with a polarization angle $\theta$ with respect to the $x$ crystallographic direction scales with:
\be
\lb{ktheta}
K(\theta)=K_{A_{1g}}+K_{B_{1g}}\cos^2(2\theta)
\ee
where $K_{A_{1g}/B_{1g}}=(K_{xx;xx}\pm K_{xx;yy})/2$. The resulting $I^{THG}(\theta)\propto |K(\theta)|^2$ is shown in Fig.\ \ref{fig3}d. According to Eq.\ \pref{ktens}, for JPM is $K_{A_{1g}}/K_{B_{1g}}=2$. 
This result is rather different from the theoretical expectations for other collective modes. Indeed, in a single-band model, as appropriate for cuprates, the Higgs signal has only a $A_{1g}$ component\cite{aoki_prb15,cea_prb16}. The density-fluctuations  response has a largely dominant $B_{1g}$ symmetry in the clean case\cite{cea_prb16}, but it becomes predominantly isotropic in the presence of even a weak disorder\cite{seibold_cm20}. As a consequence, the recent observation\cite{shimano_prl18} of a sizeable $B_{1g}$ component at optimal doping in Bi2212 compounds cannot be simply ascribed to these collective excitations. On the other hand, it is worth noting that the ratio $K_{A_{1g}}/K_{B_{1g}}=2$ for JPMs only holds within the phenomenological approach based on the quantum $XY$ model, where the tensor admits the structure \pref{ktens}. Indeed, within a microscopically-derived phase-only model  the interacting terms in the phase can differ from the one obtained within the $XY$ model, as discussed for the clean case in Ref.\ \cite{benfatto_prb04}.  On this view, while one expects in general an anisotropy of the non-linear JPM response, the exact value of the $K_{A_{1g}}/K_{B_{1g}}$ ratio has to be determines within a microscopic approach.

Our work establishes the theoretical framework to manipulate and detect JPMs in  layered cuprates across the superconducting phase transition. The basic underlying mechanism relies on the excitation of two plasma waves with opposite momenta by an intense field. For the out-of-plane response, we support the well-established approach based on non-linear sine-Gordon equations,\cite{machida_prl99,machida_physc00,nori_review10,cavalleri_natphys16,cavalleri_science18} adding a complete description of thermal effects and  highlighting the possibility to tune the resonant excitation of JPMs by changing the temperature. For the in-plane response we suggest the possible relevance of JPMs to explain several puzzling aspects emerging in recent measurements in different families of cuprates.\cite{shimano_prl18,kaiser_natcomm20,shimano_cm19} 
An open question remains a quantitative estimate of the signal coming from the JPMs, as compared to the one due to Higgs or charge-modulated density fluctuations. Indeed, as the recent theoretical work done in the context of conventional superconductors demonstrated\cite{silaev_prb19,shimano_prb19,shimano_review19,seibold_cm20}, even weak disorder becomes crucial to make such a quantitative estimate, and to establish the polarization dependence of the response\cite{seibold_cm20}. Here we notice that the large value of the in-plane plasma frequency comes along with a large value for the in-plane stiffness $J_\pll$, which controls the non-linear coupling of the JPM to the e.m.\ field.  This suggest that especially near optimal doping, where $J_\pll$ attains its maximum value, a two-plasmon THG signal can be comparable to other effects. 
%Even though the large value of the in-plane plasma mode prohibits its resonant excitation, we nonetheless expect that the larger value of the in-plane superfluid stiffness can explain a sizeable role of two-JPM excitations below $T_c$. An open question remains a microscopic comparative estimate of the relative efficiency of the light pulse to excite JPM with respect to other collective excitations, like the Higgs mode and density fluctuations. 
On this perspective, the theoretical and experimental investigation of non-linear phenomena induced by intense THz pulses represents a privileged knob to probe relative strength of pairing and phase degrees of freedom in  unconventional superconducting cuprates.

\vspace{1cm} {\bf Authors contributions} F.G. and M.U. contributed equally to this work. L.B conceived the project and supervised its development. F.G. M.U. and L.B. performed the analytical calculations. F.G. and M.U. performed the numerical simulations. L.B. wrote the manuscript with inputs from all the authors. 

\vspace{1cm} {\bf Acknowledgments}
We acknowledge useful discussions with C. Castellani, A. Cavalleri, and D. Nicoletti. We thank the authors of Ref.\ \cite{cavalleri_science18} for providing us with the experimental data of $J_{\perp}(T)$ used in Fig.\ \ref{fig2}. This work has been supported by the Italian MAECI under the Italian-India collaborative project SUPERTOP-PGR04879, by the Italian MIUR
project PRIN 2017 No. 2017Z8TS5B, and by Regione Lazio (L. R. 13/08) under project SIMAP.

\bibliography{Literature.bib} 

%\newpage

\pagebreak
\clearpage

\onecolumngrid

%%%%%%%%%% Merge with supplemental materials %%%%%%%%%%
%%%%%%%%%% Prefix a "S" to all equations, figures, tables and reset the counter %%%%%%%%%%
\setcounter{equation}{0}
\setcounter{figure}{0}
\setcounter{table}{0}
\setcounter{page}{1}
\makeatletter
\renewcommand{\thesection}{S\arabic{section}}
\renewcommand{\theequation}{S\arabic{equation}}
\renewcommand{\thefigure}{S\arabic{figure}}
%\renewcommand{\bibnumfmt}[1]{[S#1]}
%\renewcommand{\citenumfont}[1]{S#1}
%%%%%%%%%% Prefix a "S" to all equations, figures, tables and reset the counter %%%%%%%%%%

\section*{\large{S\lowercase{upplementary} I\lowercase{nformation}}\\ \vspace{0.5cm}
N\lowercase{on-linear} T\lowercase{erahertz} d\lowercase{riving of} p\lowercase{lasma} w\lowercase{aves in} l\lowercase{ayered} c\lowercase{uprates}}
\vspace{-0.05cm}
\begin{center}
Francesco Gabriele,$^1$ Mattia Udina,$^1$ and Lara Benfatto$^{1,*}$\\ \vspace{0.1cm}
\small{\emph{$^1$Department of Physics and ISC-CNR, ``Sapienza'' University of Rome, P.le A. Moro 5, 00185 Rome, Italy}}
\end{center}

\section{Effective quantum action and analogy with the non-linear sine-Gordon equations for the Josephson plasma mode} 
The derivation of the quantum action for the phase degrees of freedom can be done following a rather standard approach, see e.g.\ Ref.s \cite{nagaosa,benfatto_prb01,benfatto_prb04} and references therein. The basic formalism relies on the quantum-action representation of a microscopic superconducting model in the presence of long-range Coulomb interactions. The collective variables corresponding to the amplitude, phase and density degrees of freedom are introduced via an Hubbard-Stratonovich decoupling of the interacting superconducting and Coulomb term. This allows one to integrate out explicitly the fermionic degrees of freedom in order to obtain a quantum action in the collective-variables only, whose coefficients are expressed in terms of fermionic susceptibilities, computed on the SC ground state. The result for the Gaussian phase-only action in the isotropic three-dimensional case reads:%
\be
\lb{sgauss1}
S_{eff}[\theta]=\frac{1}{8}\sum_{i\omega_m,\bq} \left[ {\hbar^2\omega_m^2}\tilde\chi_{\rho\rho}+{D_s}\bq^2\right]|\phi(i\omega_m,\bq)|^2.
\ee
Here $D_s= \hbar^2c^2/4\pi e^2\lambda^2$ and $\tilde \chi_{\rho\rho}$ is the density-density susceptibility dressed at RPA level by the Coulomb interaction $V(\bq)$:
\be
\lb{rpa}
\tilde \chi_{\rho\rho}=\frac{\chi_{\rho\rho}^0}{1+V(\bq)\chi_{\rho\rho}^0},
\ee
where $\chi_{\rho\rho}^0$ represents the bare charge susceptibility, which reduces in the static limit to the compressibility of the electron gas, i.e.\ $\chi^0_{\rho\rho}(\omega_m=0,\bq\ra 0)\equiv \kappa$. The nature of the Goldstone phase mode is dictated by the form of the charge susceptibility. For the neutral system Coulomb interactions are absent and $\tilde \chi_{\rho\rho}$ in Eq.\ \pref{sgauss1} can be replaced by the bare one $\chi_{\rho\rho}^0$. Thus, in the long-wavelength limit the pole of the Gaussian phase propagator defines, after analytical continuation to real frequencies $i\omega_m\ra \omega+i\delta$, a sound-like Goldstone mode: 
$\omega^2=(D_s/\kappa)\bq^2$. On the other hand, in the presence of Coulomb interaction the long-wavelength limit of the charge compressibility \pref{rpa} scales as $\tilde\chi_{\rho\rho}\ra 1/V(\bq)$. In the usual isotropic three-dimensional case $V(\bq)=4\pi e^2/\bq^2$, where $\e$ is the  background dielectric constant, and one easily recovers from Eq.\ \pref{sgauss1} that 
\bea
\lb{scharged}
S_{eff}[\phi]
&=&\frac{1}{2}\sum_{i\omega_m,\bq}  \frac{\hbar^2}{4V(\bq)}\left[ \omega_m^2+\omega_P^2\right]|\phi(i\omega_m,\bq)|^2,
\eea
where $\omega_P^2\equiv 4\pi e^2 D_s/\hbar^2 \e=c^2/\l^2\e$ coincides with the   usual 3D plasma frequency. In the case of cuprates one should start from a layered model where the in-plane and out-of-plane superfluid densities are anisotropic, so that the  $D_s\bq^2$ term in Eq.\ \pref{sgauss1} is replaced by $(4D_\perp/d^2) \sin^2(k_zd/2)+D_\pll k_\pll^2$, with $D_{\perp/\pll}= \hbar^2c^2/4\pi e^2\lambda_{c/ac}^2$. In addition, one can also introduce an anisotropic expression for the Coulomb interaction, to account for the discretization along the $z$ direction\cite{benfatto_prb01}. Following e.g.\ the derivation of Ref.\ \cite{benfatto_prb01} one then recovers in the long-wavelength limit the two expressions \pref{sgaussapp} and \pref{sgaussin} in the main text. 

%Notice that at long-wavelengths the result \pref{sgaussapp} coincides also with the one based  on the non-linear sine-Gordon equations, as shown in Ref.\ \cite{machida_prl99,machida_physc00,nori_review10}. In this case however the effect of long-range forces is included via the coupling to the electromagnetic gauge and scalar potentials, which are elimiated to derive the equations of motion for the phase variables.

%For what concerns the in-plane phase mode one can follow a similar derivation. In this case, the main difference if that the phase gradient in the plane can occur in general along both $(x,y)$ crystallographic directions, and that the in-plane plasma mode is much larger than the out-of-plane one\cite{shibauchi_prl94,panagopoulos_prb96,bonn_prl04}. The final expression \pref{sgaussin} coincides with the in-plane phase-only action derived e.g. in Ref.\ \cite{benfatto_prb01,benfatto_prb04} within the context of cuprates by means of an interacting microscopic model for layered cuprates.

An alternative but equivalent approach is instead the one followed e.g.\ in Ref.s \cite{machida_prl99,machida_physc00,nori_review10,cavalleri_natphys16,cavalleri_science18}, where one deals with the  equations of motion for the plasmon, coupled to the electromagnetic fields. The connection between the two approaches has been derived in details in Ref.\ \cite{machida_prl99,machida_physc00}. Once more, the authors start from a microscopic SC layered model, and integrate out the fermionic degrees of freedom in order to build up an effective action for the phase field. The effective quantum action then reads:
\be
\lb{Sxy}
S=\sum_{n,\br_i}  \int d\tau \left\{ \frac{C_0}{2}\left(\frac{1}{2e}\frac{\pd \phi_n}{\pd \tau}\right)^2
-J_\perp \cos(\phi_{n,\br_i}(\tau)-\phi_{n+1,\br_i}(\tau))-\sum_{\a=x,y} J_\pll \cos(\phi_{n,\br_i}(\tau)-\phi_{n,\br_i+\hat\delta_\alpha}(\tau))\right\},
\ee
where $\br_i$ is the two-dimensional in-plane coordinate running over each SC layer and $\delta_\alpha$ is the in-plane versor along the direction $\a=x,y$. In Eq.\ \pref{Sxy} the quantum term accounts for the capacitive coupling $C_0=s/4\pi R_D^2$ between the planes, $s$ and $R_D$ being, respectively, the layer thickness and the Debye length. By retaining leading orders {in $\phi$} in the cosine terms the Gaussian action of Eq.\ \pref{Sxy} describes a sound mode, in full analogy with Eq.\ (\ref{sgauss1}) in the absence of RPA resummation of the density response. Indeed, as emphasized in the main text, the presence of long-range interactions is crucial in order to lift the sound mode to a plasmon. In Ref.\ \cite{machida_prl99,machida_physc00} this is achieved by adding explicitly the electric and magnetic fields, and the corresponding scalar and vector potentials. To describe the out-of-plane JPM one needs an electric field $\tilde{E}^z_{n,n+1}$ polarized perpendicularly to the planes. The magnetic field will then lie in the plane, and we can take without loss of generality $\tilde{B}^y_{n,n+1}$ along the $y$ in-plane direction. Hence the Gaussian action becomes:
%
%\begin{widetext}
\be
\lb{full}
S=\sum_{n,\br_i} \int d\tau \left\{ \frac{C_0}{2}\left(\frac{1}{2e}\frac{\pd \phi_n}{\pd \tau}+\tilde{V}_n\right)^2
+\frac{J_\perp}{2} (\phi_n-\phi_{n+1})^2+
\frac{J_\pll}{2}\left(\Delta_x \phi_n-2e a\tilde{A}^x_n\right)^2+\frac{1}{8\pi D}\left[(\e \tilde{E}^z_{n,n+1})^2+(\tilde{B}^y_{n,n+1})^2\right]\right\},
\ee
%\end{widetext}
%
where $\Delta_x\phi_n\equiv \phi_{n,\br_i}(\tau)-\phi_{n,\br_i+\hat x}$, $a$ and  $D=d+s\simeq d$ are, respectively, the in-plane and out-of-plane lattice spacings. By means of the Maxwell equations one can replace  $\tilde{E}^z_{n,n+1}=(\tilde{V}_n-\tilde{V}_{n+1})/D$ and $\tilde{B}^y_{n,n+1}=(\tilde{A}^x_{n+1}-\tilde{A}^x_{n})/D$ into Eq.\ \pref{full}. The explicit integration of the  e.m. potentials then leads:
\be
\lb{sgauss}
S^G_\perp=\frac{C}{2}\sum_{i\omega_m,\bq}  \frac{4\sin^2(k_z d/2)}{1+4\a \sin^2(k_z d/2)}\left[ \omega_m^2+\omega_P^2(\bq)\right]
|\phi(i\omega_m,\bq)|^2,
\ee
where $\a=C/C_0$ and 
\be
\lb{ompq}
\frac{\omega^2_P(\bq)}{\omega_J^2}=1+4\a \sin^2(k_z d/2)+\frac{4\left({\l_c}/{\xi_0}\right)^2 \sin^2\left({k\xi_0}/{2}\right)}{1+4(\l_{ab}/d)^2 \sin^2(k_z d/2)}
\ee
describes the full dispersion of the plasma mode as a function of $\bq=(k_z,k)$, with $k$ laying in the in-plane propagation direction. The pole equation for the Gaussian phase mode, i.e. $\omega^2=\omega_P^2(\bq)$, is completely equivalent to the solution of the  linearized sine-Gordon equation for Josephson plasma waves previously addressed in the literature\cite{machida_prl99,machida_physc00,nori_review10,cavalleri_natphys16,cavalleri_science18}. In cuprates the constant $\a$ is usually very small, so the main dispersion of the plasmon comes from the last term of Eq.\ \pref{ompq}, which accounts for the inductive coupling between planes. In this approximation, the Gaussian phase fluctuations identify a collective mode whose energy dispersion is obtained as the pole of the Guassian propagator for phase fluctuations:
\be
\lb{prop}
\langle|\phi(i\omega_m,\bq)|^2 \rangle=\frac{1}{4\sin^2(k_z d/2)\left[ \omega_m^2+\omega_P^2(\bq)\right]}.
\ee
By analytical continuation $i\omega_m\ra \omega+i\delta$ in Eq.\ \pref{prop} we then get 
\be
\lb{disp}
\omega^2=\omega^2_P(\bq)=\omega_J^2 \left[1+\frac{4\left({\l_c}/{\xi_0}\right)^2 \sin^2\left({k\xi_0}/{2}\right)}{1+(\l_{ab}/d)^24 \sin^2(k_z d/2)}\right].
\ee
The relation \pref{disp} is the same that one obtains by using the equation of motion approach discussed in Ref.s\ \cite{machida_prl99,machida_physc00,nori_review10,cavalleri_natphys16,cavalleri_science18}. In this case, one introduces directly the variable $\theta_n \equiv \phi_n-\phi_{n+1}$ which represents the phase difference between nearest-neighbour layers. It is then shown to satisfy the equation of motion\cite{nori_review10}
\be
\lb{SineGordon}
\left(1-\frac{\l_{ab}^2}{d^2}\pd^2_n\right)\left[\frac{1}{\omega_J^2}\frac{\partial^2\theta_n}{\partial {t}^2}+\sin(\theta_n)\right]-\frac{\l_c^2}{\xi_0^2}\partial_x^2\theta_n=0,
\ee
where $\pd_n^2 f_n\equiv f_{n+1}+f_{n-1}-2f_n$ is the second-order discrete differential operator along the $z$ direction, and analogously $\pd^2_x$ for the $x$ direction. As one can easily check, when $\sin \theta_n \approx \theta_n$ Eq.\ \pref{SineGordon} admits a wave solution $\theta_n(x=m\xi_0,t)\propto\exp [i(kx+k_z nd-\omega t)]$  where the frequency $\omega$ and the momentum $\bq=(k,k_z)$ satisfy Eq.\ \pref{disp}. In the  approach of Ref.s \cite{machida_prl99,machida_physc00,nori_review10,cavalleri_natphys16,cavalleri_science18}, based on the study of the equation of motions,  the electromagnetic field is completely eliminated and the non-linear effects are included by retaining the full $\sin\theta_n$  term in the sine-Gordon model \pref{SineGordon}. In this case, a real $\bq$ solution for a propagating waves is only possible if one retains the full momentum dispersion in Eq.\ \pref{ompq}.
In contrast, in our approach the plasma mode is first computed at Gaussian level, and then non-linear effects originate by 
retaining in Eq.\ \pref{full} the full cosine term appearing in Eq.\ \pref{Sxy}, which is responsible for non-linear coupling to the gauge potential. The phase mode is then integrated out in order to obtain the complete electromagnetic response, as required to describe non-linear effects in the currents, see Eq.\ (3) and (6) in the main text. This is indeed the same approach that has been used so far to investigate the nature and the non-linear response of the SC Higgs mode, by means of two-dimensional models able to describe the in-plane response\cite{cea_prb16}.
In this view, the dispersion of the plasma mode (as well as the Higgs mode in the case of amplitude fluctuations) is quantitatively irrelevant, and what matters is only the resonance process which occurs when the pumping frequency matches the value of the Josephson frequency $\omega_J$. For this reason we retained in our calculation the $\bq \ra 0$ long-wavelength limit in the phase propagator \pref{prop}, as done in Eq.\ (4) of the main text. 

To account also for possible dissipative effects one usually adds in the equation \pref{SineGordon} also a term linear in the time derivative:
\be
\lb{SineGordondiss}
\left(1-\frac{\l_{ab}^2}{d^2}\pd^2_n\right)\left[\frac{1}{\omega_J^2}\frac{\partial^2\theta_n}{\partial {t}^2}+r\frac{\partial\theta_n}{\partial t}
+\sin(\theta_n)\right]-\frac{\l_c^2}{\xi_0^2}\partial_x^2\theta_n=0.
\ee
This additional tern can be justified once more at microscopic level by following the derivation of Ref.\ \cite{benfatto_prb01}. Indeed, the $D_s\bq^2$ term in Eq.\ (15) of the main text originates from the long-wavelength limit of the current-current correlation function. Taking into account also the presence of a regular part $\s_{reg}$ of the low-frequency conductivity due to normal quasiparticles, one can easily show\cite{benfatto_prb01} that  Eq.\ (4) of the main text gets modified as:
\be
\lb{DissAction}
S_\perp^{G,diss}=\frac{v}{2d^2}\sum_{i\omega_m,k_z} 4\sin^2(k_z d/2)\left(\omega_m^2+\omega_J^2+\s_{reg}\mid \omega_m \mid \right)\mid \phi(i\omega_m,k_z) \mid^2.
\ee
Using again the analogy between the pole of the phase propagator \pref{prop} and the solution of the equations of motion \pref{SineGordondiss} one understands why absorption by normal quasiparticles leads to dissipation of the plasma waves.

\section{Derivation of the non-linear kernel} 
The current $I_\alpha$ in the $\alpha=(x,y,z)$ direction is defined as usual through the functional derivative with respect to $A_\alpha$  of the action $S_\bA$:
\be
\lb{I_alfa}
I_\a=-\frac{\d S_\bA}{\d A_\a}.
\ee
As a general rule, to compute the third-order contribution to $I_\a$ in Eq.\ \pref{I_alfa} one needs to expand the e.m.\ action up  to fourth-order terms in $\bA$. For example the coupling term of the JPM to $A_z^2$ in Eq.\  \pref{stot} in the main text leads to a $A_z^4$ contribution after integrating out the plasmon\cite{suppl}. This is represented by the Feynmann diagram of Fig.\ \ref{fig1}b. There each solid line denote the Gaussian phase mode, obtained from Eq.\ \pref{sgaussapp} as $
\langle|\phi(i\omega_m,k_z)|^2 \rangle=\left\{ 4\sin^2(k_z d/2)\left[ \omega_m^2+\omega_J^2\right]\right\}^{-1}$. 

\begin{figure}[h!]
\centering
\includegraphics[scale=1.]{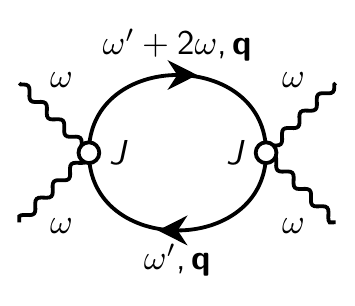}
\caption{Feynman diagram representing the fourth-order term \pref{SFourthOutPlane}. Here wavy/solid lines represent the vector potential/phase mode, respectively. The phase mode is computed at Gaussian level and it is given by Eq.\ \pref{prop}. Here $\bq=(k,k_z)$ denotes the momentum of the JPM, being $k_z$ for the out-of-plane JPM, and $k=(k_x,k_y)$.}
\label{figSdia}
\end{figure}

Let us first focus now on the out-of-plane THG. As shown in Eq.\ (5) of the main text, by expanding the first cosine term in Eq.\ (\ref{Sxy}), once the minimal-coupling $\phi_n-\phi_{n+1}\rightarrow\phi_n-\phi_{n+1}-\frac{2\p}{\Phi_0}dA_z$ has been performed, we find that:
\bea
\lb{sexpprp}
S&=&S^G_\perp-\frac{\pi^2 d^2}{\Phi_0^2}
\sum_{n,\br_i} J_\perp \int d\tau A_z^2( \tau)\left( \phi_{n,\br_i}(\tau)-\phi_{n+1,\br_i}(\tau) \right)^2+...= \nn\\
&=&S^G_\perp -\sqrt{\frac{T}{N_s}}\frac{\pi^2 d^4}{\Phi_0^2}\sum_{i\omega_m,i\omega_m'}\sum_{k_z} J_\perp k_z^2 A_z^2(i\omega_m-i\omega_m')\overline{\phi}(i\omega_m,k_z)\phi(i\omega_m',k_z)+...
\eea
where dots denote additional terms not relevant for the $\chi^{(3)}$ response, and we used the $k_z\ra 0$ limit for the finite difference along $z$. Starting from Eq.\ \pref{sexpprp} we can easily obtain the out-of-plane non-linear optical kernel. After adopting the notation $q\equiv (i\omega_m,k_z)$ the partition function of the system can be rewritten as:
\be
\lb{PartFun}
Z=\int \mathcal{D}[\phi,\overline{\phi}] \text{e}^{-\frac{1}{2}\sum_{q,q'}\left[D(q,q')+R(q,q')\right]\overline{\phi}(q)\phi(q')},
\ee
where: 
\bea
\lb{Dqq'Rqq'}
&D&(q,q')=v\left(\omega_m^2+\omega_J^2\right)k_z^2 \d(i\omega_m-i\omega_m')\d(k_z-k_z') \\
\lb{rqq}
&R&(q,q')=-\sqrt{\frac{T}{N_s}}\frac{2 \pi^2 d^4}{\Phi_0^2} J_\perp k_z^2 A_z^2(i\omega_m-i\omega_m') \d(k_z-k_z').
\eea
At this point we integrate out the phase fluctuations: this gives a contribution beyond RPA approximation to the effective action equivalent to one-loop corrections in the phase mode. With straightforward algebra we obtain:
%In analogy with ordinary gaussian integrals we find that:
%%
%\be
%\lb{GaussInt}
%\int \mathcal{D}[\theta,\overline{\theta}] \text{e}^{-\frac{1}{2}\sum_{q,q'}\left(D(q,q')+R(q,q')\right)} \propto \frac{1}{\det_q[D+R]}
%\ee
%%
%$\det_q$ being the determinant on the $q$'s degrees of freedom. We can express the determinant as a power series by means of the Binet formula ($\det[AB]=\det[A]\det[B]$) and of the Taylor expansion of the logarithmic function $\ln(1+x)=\sum_{n=1}^{+\infty}\frac{x^n}{n}$, indeed:
%%
%\bea
%\lb{expDet}
%\det_q[D+R]&=&\det_q[D(1+D^{-1}R)]=\det_q[D]\det_q[1+D^{-1}R]=\\
%&=&\det_q[D]\text{e}^{\text{tr}_q\{\ln(1+D^{-1}R)\}}=\det_q[D]\text{e}^{\sum_{n=1}^{+\infty}\frac{\text{tr}_q\{(D^{-1}R)^n\}}{n}}
%\eea
%%
%where $\text{tr}_q$ is the trace on the $q$'s degrees of freedom, and we also used the fact that $\det[A]=\text{e}^{\text{tr}\{A\}}$. The effective action becomes, therefore:
%
\be
\lb{seff}
S^{(eff)}=\sum_{n=1}^{+\infty}\frac{\text{Tr}_q\{(D^{-1}R)^n\}}{n}.
\ee
%
%We dropped the term $\frac{1}{\det_q[D]}$ of the partition function, since it is a multiplicative constant which gives no contributions to the THG. 
Therefore, the fourth-order action in $\mathbf{A}$ $S^{(4)}$ is:
\be
S^{(4)}_\perp =K_0 J_\perp^2 \sum_{i\omega_m} A_z^2(i\omega_m) T\sum_{i\omega_m'} \frac{ 1 }{[(\omega_m+\omega_m')^2+\omega_J^2][(\omega_m')^2+\omega_J^2]} A_z^2(-i\omega_m)
\label{SFourthBeforeSummation}
\ee
where we retained the only term contributing to the THG and we put all the multiplicative constants into $K_0$. Eq.\ \pref{SFourthBeforeSummation} corresponds to the Feynman diagram of Fig.\ 1b of the main text, that we reported in Fig.\ \ref{figSdia} with the explicit frequency dependence. Here each solid line denotes a phase propagator, given by Eq.\ \pref{prop} above. Notice that since the gauge field couples to the phase gradient along $z$, see Eq.\ \pref{rqq}, each $k_z^2$ term from the vertex of the diagram in Fig.\ \ref{figSdia} compensates a $1/k_z^2$ term from the denominator of the phase propagator in Eq.\ \pref{prop}. 
After computing the Matsubara sum, i.e.\ $T\sum_{i\omega_m'}\frac{1}{(\omega_m+\omega_m')^2+\omega_J^2}\frac{1}{i\omega_m^2+\omega_J^2}=\frac{\coth\left(\frac{\b\omega_J}{2}\right)}{\omega_J(\omega_m^2+4\omega_J^2)}$, the fourth-order effective action becomes:
\be
S^{(4)} = \sum_{i\omega_m} A_z^2(i\omega_m) K_\perp (i\omega_m) A_z^2(-i\omega_m),
\label{SFourthOutPlane}
\ee
where
\be
\lb{kperp}
K_\perp(i\omega_m)=K_0\frac{J_\perp^2}{\omega_J}\frac{\coth(\beta \omega_J/2)}{4\omega_J^2+\omega_m^2}
\ee
is the out-of-plane non-linear optical kernel.

Let us consider now the case of an in-plane polarized external e.m.\ field. Following the same scheme adopted for the out-of-plane case we find that the in-plane fourth-order effective action is:
\be
S^{(4)} = \sum_{i\omega_m} \sum_{i,j} A_i^2(i\omega_m) K_{ij} (i\omega_m) A_j^2(-i\omega_m),
\label{SFourthInPlane}
\ee
where $K_{ij}(i\omega_m)=M_{ij}\frac{J_\pll^2}{\omega_J^{\pll}}\frac{\coth(\beta \omega_J^{\pll}/2)}{4(\omega_J^{\pll})^2+\omega_m^2}$. $M_{ij}=K_0\begin{pmatrix} \sum_{k_x,k_y}\frac{k_x^4}{\mathbf{k}^4} & \sum_{k_x,k_y}\frac{k_x^2 k_y^2}{\mathbf{k}^4} \\
\sum_{k_x,k_y}\frac{k_x^2 k_y^2}{\mathbf{k}^4} & \sum_{k_x,k_y}\frac{k_x^4}{\mathbf{k}^4} \end{pmatrix}$ is the polarization-dependent tensor, whose components read:
\be
M_{xx}=M_{yy} \simeq \int_{0}^{2\pi} d\phi \cos^4 \phi = \frac{3\pi}{4}
\label{Mxx}
\ee
\be
M_{xy}=M_{yx} \simeq \int_{0}^{2\pi} d\phi \cos^2 \phi \sin^2 \phi = \frac{\pi}{4}.
\label{Mxy}
\ee
Hence the tensor components of the in-plane non-linear optical kernel are those enlisted in Eq.\ (12) of the main text. 
%provided that we put $K_\pll \equiv \frac{\p}{4}K_0$. 
%If $\theta$ is the polarization angle of the pump field, i.e. $\mathbf{A}=(A\cos\theta,A\sin\theta,0)$, then to current in the same direction of the applied field is proportional to the in-plane non-linear optical kernel:
%%
%\be
%K(\theta)=K_{xx}(\cos^4\theta+\sin^4\theta)+2K_{xy}\cos^2\theta\sin^2\theta
%\label{Ktheta2}
%\ee
%%
%Eq. ($\ref{Ktheta2}$) can be easily rewritten in the form of Eq. ($\ref{ktheta})$.\\

%\section{T-dependence of the out-of-plane and in-plane superfluid stiffness}

%
%\begin{figure}[h!]
%\centering
%\includegraphics[scale=0.6]{figS2_cav.pdf}
%\caption{{\bf Out-of-plane and in-plane superfluid densities} (a)-(b) $T$-%%dependence of the superfluid densities.}
%\end{figure}
%
As a final step, let us show how the additional $\s_{reg}|\omega_m|$ term of Eq.\ \pref{DissAction} can be added to the non-linear kernel in order to account for the effect of dissipation. In this case, the calculation is done by  introducing a finite spectral function to the phase mode $A(z)\equiv \frac{z \sigma}{ \left( z^2 - \omega_J^2 \right)^2 + z^2 \sigma_{reg}^2 }$. One then finds that, in general, the kernel becomes: 
\be
\lb{Kdiss}
K^{(diss)}(i\omega_m)=K_0 J_\perp^2
\int_{-\infty}^{+\infty} \frac{dz}{\pi} 
\int_{-\infty}^{+\infty} \frac{dz'}{\pi}
A(z) A(z') \frac{ b(z) - b(z') }{ z'-z + i\omega_m },
\ee
where $b(z)=\frac{1}{e^{\beta z} - 1 }$ is the Bose function. If $\s<<\omega_J$ it can be shown that Eq.\ \pref{Kdiss} can be approximated, after analytical continuation, as:
\be
\lb{KDissApprox}
K^{(diss)}(\omega)\simeq K_0  
\frac{J_\perp^2}{\omega_J}
\frac{ \coth\left(\frac{\beta\omega_J}{2}\right) }
{4\omega_J^2 - \left( \omega + i\sigma_{reg} \right)^2}.
\ee
Eq.\ (\ref{KDissApprox}) is the expression used, indeed, to compute all the quantities of interest in the main text. In analogy with Ref.\ \cite{nori_review10} we also assumed that 
\be
\lb{SigmaR}
\sigma_{reg}=\g(T) \equiv \g_0+r(T), 
\ee
where $r(T)=r_0 e^{-\Delta/T}$\cite{nori_review10}, and $\g_0$ is a small regularization constant, which prevents the non-linear optical kernel $K^{(diss)}$ to be ill-defined at $T=0$. Both $\g_0$ and $r_0$ parameters are fixed by looking at the number of time-resolved oscillations observed experimentally in the pump-probe set up of Ref.\ \cite{cavalleri_natphys16} at low temperatures. To better reproduce the experimental findings, in Fig.\ \ref{fig2} of the main text we fixed $\gamma_0/2\pi=0.08$ THz, while $r_0=0.3 \omega_{J,0}$ in panels c,d and $r_0=0.6 \omega_{J,0}$ in panels e-h. There $\omega_{J,0}/2\pi=0.47$ THz is the out-of-plane plasma frequency at $T=0$. In Fig.\ \ref{fig3}, instead, we set  $\gamma_0=0.1\omega_{J,0}$, where now $\omega_{J,0}/2\pi = 240$ THz is the $T=0$ value of the in-plane plasma frequency.

\section{Modelling of the Broad-band pump pulse}
For a narrow-band multicycle pulse one can assume a monochromatic incident field, and the THG is simply related to the non-linear optical kernel via Eq.\ \pref{ithg}. However, for a  broad-band pulse with central frequency $\Omega$, the THG is more generally associated with the $3\Omega$ component in the nonlinear current\cite{cea_prb16,udina_prb19}:
\be
I^{NL}_i(\omega) = \sum_j \int d\omega' A_i(\omega-\omega')K_{ij}(\omega')A_j^2(\omega'),
\label{nonlin}
\ee
as shown e.g.\ in Fig.\ \ref{fig2}e in the main text (with $i,j=z$ and $K_{ij}=K_\perp$) at different temperatures. Here, $A_z(\omega)$ is given by the Fourier transform of $A_z(t)=A_0e^{-t^2/\tau^2}\sin{(2\pi \Omega t)}$, while $A_z^2(\omega)$ is defined as the Fourier transform of $A_z^2(t)$. The $\tau=0.85$ ps and  $\Omega / 2\p=0.45$ THz parameters are set in such a way that the e.m.\ field $E_z(t) \propto -\partial A_z(t)/ \partial t$ well reproduces the experimental pulse profile of Ref.\ \cite{cavalleri_science18}.

\section{ Pump-probe configuration}
In a pump-probe experiment designed to excite the out-of-plane JPM both the pump and probe fields are polarized along $z$, i.e. $E_z=E_{probe}(t)+E_{pump}(t)$. Here we will refer for simplicity to the transmission configuration, as discussed in Ref.\ \cite{giorgianni_natphys19,udina_prb19}, where one measures the variation $\delta E_{probe}(t)$ of the transmitted probe field with and without the pump, so that terms not explicitly depending on the pump field cancel out. This allows one to express it as  $\delta E_{probe}(t) \propto \int dt' A^{probe}_z(t)K(t-t')(A^{pump}_z)^2(t')$. By considering a fixed $t_g$ acquisition time and implementing the time-delay $t_{pp}$ between the pump and the probe,  $\delta E_{probe}(t_g;t_{pp})$ becomes a function of $t_{pp}$ only, as given by the first line of Eq.\ \pref{osc}. Finally, by computing from Eq.\ \pref{knonl} the non-linear kernel in time domain, i.e. $K(t)=\int\frac{d\omega}{2\p}K(\omega)e^{-i\omega t}=F(T)e^{-\g t}\sin(2\omega_J t)$, we derive the last line of Eq.\ \pref{osc}.  

For the reflection geometry used in Ref.\ \cite{cavalleri_natphys16} the basic mechanism is the same, so that one expects that  the differential reflectivity signal scales with the convolution of the non-linear kernel times the pump field squared given in Eq.\ \pref{osc}. For the calculation of Fig.\ \ref{fig2}h in the main text we used the simulation of the broad-band pump field explained above. 
 For the in-plane response measured in Ref.\ \cite{shimano_cm19}, the huge frequency mismatch between the spectral components of the gauge field and $2\omega_J^\pll$  implies that only the term with $t=t_{pp}$ survives in the integral \pref{osc}. As a consequences the oscillations are absent and $\delta E_{probe}(t_{pp})$ simply scales as the square of the pump field, modulated by $F(T)$ and by the polarization encoded in the kernel \pref{ktens}. Indeed, if the pump field forms an angle $\theta$ with the $x$ axis and the probe is applied e.g.\ along the $x$ axis, from Eq.\ \pref{osc}, properly generalized for the planar configuration, one easily sees that $\delta E_x\sim K_{xx;xx}\cos^2\theta+K_{xx;yy}\sin^2\theta=K_{A_{1g}}+K_{B_{1g}}\cos(2\theta)$. This is exactly the decomposition used to analized the transient reflectivity measured in Ref.\ \cite{shimano_prl18}.

\end{document}